\documentclass[prb,aps,twocolumn,superscriptaddress,floatfix,showpacs,longbibliography]{revtex4-2}
\usepackage{graphicx}
\usepackage{dcolumn}
\usepackage{bm}
\usepackage{upgreek}
\usepackage{natbib}
\usepackage[normalem]{ulem}
\usepackage{amsmath,amssymb}
\usepackage{dsfont}
\usepackage[english]{babel}
\usepackage{braket}
\usepackage{diagbox}
\usepackage{color}

\begin{document}
\title{Topological exciton bands and many-body exciton phases in transition metal dichalcogenide trilayer heterostructures}
\author{Ze-Hong Guo}
\thanks{These authors contributed equally to this work.}
\affiliation{Guangdong Basic Research Center of Excellence for Structure and Fundamental Interactions of Matter, Guangdong Provincial Key Laboratory of Quantum Engineering and Quantum Materials, School of Physics, South China Normal University, Guangzhou 510006, China}

\author{Tao Yan}
\thanks{These authors contributed equally to this work.}
\affiliation{Guangdong Basic Research Center of Excellence for Structure and Fundamental Interactions of Matter, Guangdong Provincial Key Laboratory of Quantum Engineering and Quantum Materials, School of Physics, South China Normal University, Guangzhou 510006, China}

\author{Jin-Zhu Zhao}
\affiliation{Guangdong Basic Research Center of Excellence for Structure and Fundamental Interactions of Matter, Guangdong Provincial Key Laboratory of Quantum Engineering and Quantum Materials, School of Physics, South China Normal University, Guangzhou 510006, China}
\affiliation{Guangdong-Hong Kong Joint Laboratory of Quantum Matter, Frontier
Research Institute for Physics, South China Normal University, Guangzhou
510006, China}

\author{Yuan-Jun Jin}
\affiliation{Guangdong Basic Research Center of Excellence for Structure and Fundamental Interactions of Matter, Guangdong Provincial Key Laboratory of Quantum Engineering and Quantum Materials, School of Physics, South China Normal University, Guangzhou 510006, China}
\affiliation{Guangdong-Hong Kong Joint Laboratory of Quantum Matter, Frontier
Research Institute for Physics, South China Normal University, Guangzhou
510006, China}

\author{Qizhong Zhu}
\email{qzzhu@m.scnu.edu.cn}
\affiliation{Guangdong Basic Research Center of Excellence for Structure and Fundamental Interactions of Matter, Guangdong Provincial Key Laboratory of Quantum Engineering and Quantum Materials, School of Physics, South China Normal University, Guangzhou 510006, China}
\affiliation{Guangdong-Hong Kong Joint Laboratory of Quantum Matter, Frontier
Research Institute for Physics, South China Normal University, Guangzhou
510006, China}

\date{\today}

\begin{abstract}
Twisted multilayer transition metal dichalcogenides (TMDs) are a promising platform for realizing topological exciton phases. Here we propose that twisted TMD heterotrilayers WX$_2$/MX$_2$/WX$_2$ with layer symmetry represents a realistic system for realizing topological exciton bands and interesting many-body excitonic phases, simply by tuning the twist angle. These symmetric heterotrilayers form a type-II band alignment, where the electrons are confined in the middle layer and holes are distributed among the outer two layers, for the lowest energy excitons. The outer two layers are then rotated at different centers by opposite angles, forming a helical structure. Interlayer excitons with opposite dipoles are hybridized by the coupling between outer two layers, resulting in topological moir\'e exciton bands. Furthermore, by constructing a three-orbital tight-binding model, we map the many-body phase diagram of interacting dipolar and quadrupolar excitons at different twist angles and exciton densities and reveal the existence of sublattice-dependent staggered superfluid and Mott insulator phases. The recent experimental observation of quadrupolar excitons in symmetric heterotrilayers brings the intriguing phases predicted in this study within immediate experimental reach.   
\end{abstract}

\maketitle

\section{introduction}

Twisted two-dimensional (2D) materials such as graphene and transition metal dichalcogenides (TMDs) have become a powerful platform for realizing a variety of topological and strongly correlated phases \cite{kennes2021moire}. In particular, an extraordinary breakthrough is the proposal \cite{wu2019topological,yu2020giant,devakul2021magic,li2021spontaneous,crepel2023anomalous} and realization of  fractional quantum anomalous Hall insulator without the need of external magnetic field \cite{zeng2023thermodynamic,xu2023observation,park2023observation,cai2023signatures,lu2024fractional}. Even more interestingly, there is experimental evidence of fractional quantum spin Hall insulator \cite{kang2024evidence}. The breakthrough not only provides an unprecedented platform for exploring fundamental theories related to fractional topological phases, but also potentially enables practical applications such as topological quantum computation based on non-Abelian anyons \cite{moore1991nonabelions,wen1991non,nayak2008non}. 

Inspired by the progress in realizing correlated and topological electronic states, twisted 2D materials also provide an appealing system for realizing various bosonic many-body phases. In particular, due to the presence of moir\'e potential formed by the relative twist between TMD bilayers, the charge carriers separated to different layers are trapped in a series of moir\'e potential minima, resulting in the formation of moir\'e excitons \cite{yu2017moire,jin2019observation,forg2021moire,tran2019evidence,seyler2019signatures,tran2019evidence,alexeev2019resonantly,liu2021signatures}. The long lifetime of interlayer moir\'e excitons enable the achievement of exciton many-body phases at equilibrium, such as the exciton Mott insulator phase recently observed \cite{lagoin2022extended,lagoin2022mott,park2023dipole,lian2024valley,gao2024excitonic}. 

Furthermore, twisted materials offer a promising platform for realizing bosonic topological phases.
Bosonic topological phases, such as bosonic integer and fractional quantum Hall states \cite{senthil2013integer,regnault2013microscopic,grover2013quantum,wang2011fractional,grass2014quantum,moller2009composite,he2015bosonic,hormozi2012fractional,wu2013quantum} are even more intriguing than their fermionic counterparts. However, the realization of these bosonic topological phases remains quite limited. To date, only very few platforms have successfully demonstrated these phases,
including cold atoms \cite{leonard2023realization}, photons in circuit QED lattices \cite{wang2024realization} and excitons in moat band \cite{wang2023excitonic}. 

Notably, topological exciton bands can be engineered in twisted trilayer or multilayer materials, providing a promising pathway towards the realization of exciton quantum Hall states. Recently, a theoretical breakthrough has been made with the proposal of a twisted trilayer TMD heterostructure. The proposed structure consists of twisted homobilayers adjacent to a different monolayer and is predicted to form a topological exciton band when subjected to a vertical electric field \cite{xie2024long}. While the general concept remains broadly applicable, the specific theory relies on a continuum model similar to that of twisted homobilayers, which, however, becomes questionable as the newly added third layer introduces an asymmetry to the twisted homobilayers and breaks the symmetry required to adopt a similar continuum model.
In other words, in order to incorporate the asymmetric effect appropriately, a much more complicated continuum model should be adopted. Furthermore, an external electric field has to be applied to partially compensate the asymmetry effects. Motivated by the layer-symmetric angle-aligned trilayer heterostructures recently realized in experiment \cite{du2024new,yu2023observation,li2023quadrupolar,bai2023evidence,lian2023quadrupolar,xie2023bright}, it is possible to introduce a twisted layer-symmetric heterotrilayer to realize a topological exciton band \cite{wu2017topological,chen2017chiral,kwan2021exciton,xie2024long}. Progress along this path could lead to the development of an exciton topological flat band, where the partial filling of excitons might enable the emergence of fractional bosonic topological phases \cite{wang2011fractional}.

In this paper, based on the recently realized angle-aligned TMD heterotrilayers, we introduce the twist angle in this system and propose a layer-symmetric twisted heterotrilayer model, which can be described by a continuum model similar to that of twisted homobilayers. With twisted $\rm{WSe_2/MoSe_2/WSe_2}$ as a specific example, the outer homobilayers are rotated in opposite direction with respect to the middle layer by the same angle but at different rotation centers. The middle and outer layers form a type-II band alignment, leading to dipolar exciton with opposite dipoles at lowest energy.
Due to the layer symmetry of the model, we are able to construct a continuum model based on symmetries similar with twisted homobilayers, with parameters input from first-principles calculations. Based on the continuum model for the moir\'e valence bands, we further construct a momentum space model which describes the coupling of moir\'e excitons in moir\'e potential, and find that the resulting moir\'e exciton band can become topological for certain twist angles even in the absence of external electric field. In the parameter regime considered here, the lowest three bands consistently have a total Chern number of zero. Starting from the moiré exciton band, we construct Wannier functions localized at three high-symmetry spatial positions and develop a three-orbital tight-binding (TB) model to describe the properties of the lowest three exciton bands. We further consider the exciton dipolar and quadrupolar interactions and calculate many-body phase diagram within the mean-field Gutzwiller method for different twist angles and exciton densities. The model proposed in this work serves as a realistic system for studying topological exciton band and many-body topological exciton phases.

The structure of the rest part of the paper is organized as follows. In Sec. \ref{sec2}, we propose the twisted symmetric heterotrilayer model and calculate the interlayer dipolar exciton states in this system. In Sec. \ref{sec3}, we first construct the continuum model of the moir\'e valence bands from hybridization between outer homobilayers and then derive the moir\'e exciton band model in momentum space. Based on the model, we calculate the Berry curvature and Chern number of the moir\'e exciton bands. In Sec. \ref{sec4}, for the lowest three exciton bands with total Chern number being zero, we construct three Wannier functions and develop a three-orbital TB model to describe the properties of the lowest three exciton bands. We then consider the filling of dipolar and quadrupolar excitons, and calculate the many-body exciton phase diagrams for different twist angles and exciton densities in Sec. \ref{sec5}. Finally, we discuss the drawbacks and further improvements of this model, and conclude in Sec. \ref{sec6}.

\section{system model and excitonic state}
\label{sec2}
\begin{figure}[tbp] \centering
	\includegraphics[width=0.99\linewidth]{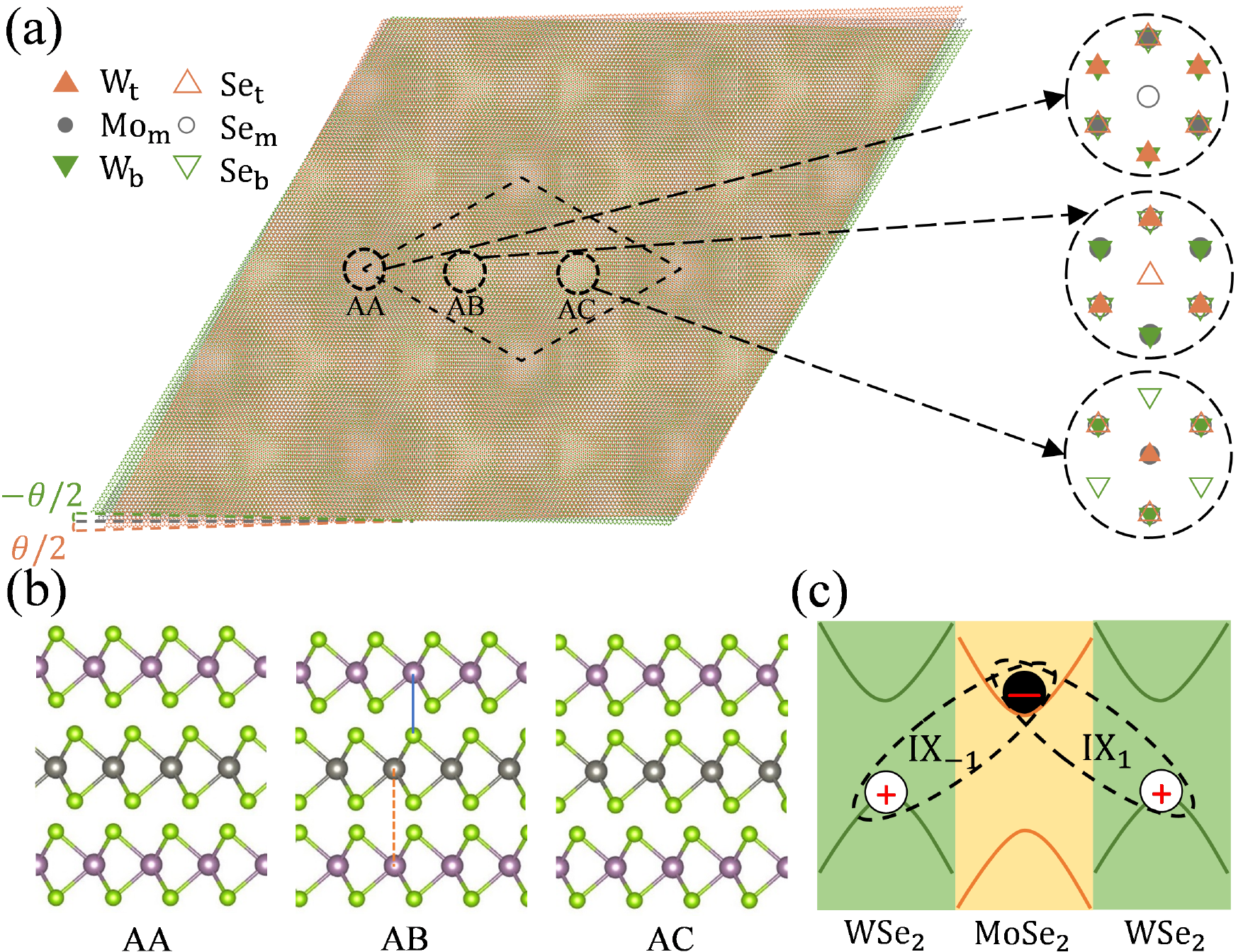}
	\caption{ (a) Schematic diagram of the proposed moir\'e heterostructures. The top (bottom) $\rm{WSe}_{\text{2}}$ layer is twisted relative to the middle monolayer $\rm{MoSe}_{\text{2}}$ by a small angle $\theta/2$ $\left(-\theta/2\right)$. (b) The side view of the three high-symmetry stacking configurations at different locations in the moir\'e unit cell shown in (a). From left to right, the AA, AB and AC stackings are displayed. 
 The solid blue and dashed red lines connecting the upper and lower bilayers represent their respective rotation centers. 
    (c) Schematic diagram of the type-II band alignment for different layers. The two types of low-energy interlayer excitons are denoted as $\rm{IX}_{1}$ and $\rm{IX}_{-1}$, with opposite dipole moments. The hybridization of $\rm{IX}_{1}$ and $\rm{IX}_{-1}$ exctions gives rise to quadrupolar excitons.}
	\label{fig1}
\end{figure}
We propose a layer-symmetric twisted TMD heterotrilayers as a realistic system to realize topological exciton band. The topological exciton bands discussed in this work focus on interlayer excitons (IXs), which possess long lifetimes that enable the realization of many-body exciton phases at the ground state, in contrast to topological intralayer exciton bands \cite{wu2018theory}. The topology here, defined in the center-of-mass momentum of IXs, differs from that of topological excitons predicted in internal degrees of freedom \cite{xie2024theory}. As illustrated in Fig. \ref{fig1}(a), we consider a heterotrilayer which is obtained by inserting an extra monolayer $\rm{MX_2}$ between a twisted homobilayer t$\rm{WX_2}$.
Here M and W represent different transition metal atoms and the common chalcogen element $\rm{X}$ can be S, Se, or Te. The choices of M, W and X atoms must ensure that the bands of middle and outer layers satisfies the type-II alignment. Such systems without twist angle have been studied in several experimental groups \cite{du2024new,yu2023observation,li2023quadrupolar,bai2023evidence,lian2023quadrupolar,xie2023bright}, and a novel type of excitons, i.e., quadrupolar excitons are observed {\cite{yu2023observation,lian2023quadrupolar}. Different with those systems, here 
the top (bottom) layer is twisted by a small angle of $+\theta/2\left(-\theta/2\right)$ relative to the middle layer at different rotation centers, resulting in the formation of long-period moir\'e superlattices between them. The rotation center for the upper bilayer is the M atom in the top layer located above the X atom in the middle layer, while the rotation center for the lower bilayer is the M atom in the middle layer positioned directly above the M atom in the bottom layer [see the AB stacking in Fig. \ref{fig1}(b)]. By neglecting the slight mismatch in lattice constants, the moir\'e period is $a_M \approx a_0/\theta$ for small twist angle.

Within the moir\'e unit cells, there are three distinct types of high-symmetry stacking configurations which are referred to as AA, AB and AC stackings (see Fig. \ref{fig1}).
The AA stacking refers to the configuration where the M atoms of top layer are aligned with the M atoms of the bottom layer, denoted as $\rm{M_t-M_b}$, the AB stacking is represented as $\rm{M_t-X_b}$, and the AC stacking corresponds to $\rm{X_t-M_b}$. 
For AA stacking, the outer two layers are mirror symmetric with respect to the middle layer. AB stacking is interchangeable with AC stacking under mirror operation. In combination with the intrinsic $\hat{C}_{3}$ rotation symmetry, these symmetry constrains dictate that the moir\'e potential due to hybridization between the outer two layers resembles that of twisted homobilayers \cite{wu2019topological,devakul2021magic}.

In the following, we focus on a specific choice of the twisted trilayer TMDs, i.e., $\rm{WSe_2/MoSe_2/WSe_2}$, where the energy bands of the three layers exhibit the type-II band alignment [see Fig. \ref{fig1}(c)] and the lattice mismatch is quite small ($<0.2\%$). Although the following calculations are done for this specific heterotrilayers, the theory is also applicable for other heterotrilayers that have type-II band alignment. With large band energy offset between the middle and outer layers, the electrons reside in conduction band of $\rm{MoSe_2}$, while the holes occupy the valence band of $\rm{WSe_2}$, for low energy excitons.
The electrons and holes within the same valley can pair though attractive Coulomb interactions, forming IXs.
IXs in $+K$ and $-K$ valley can separately form moir\'e exciton bands, and due to time-reversal symmetry, the valley Chern number of exciton band is opposite, i.e., $\mathcal{C}_{+K}=-\mathcal{C}_{-K}$. In the following discussion, we focus on the $+K$ valley moir\'e excitons and the properties of $-K$ valley moir\'e excitons can be obtained through time-reversal operation. In addition, moir\'e excitons in different valleys can be selectively excited optically.

Since the binding energy of IXs is much larger than the interlayer hopping strength \cite{yu2021luminescence},
we are allowed to solve the exciton Bethe-Salpeter equations (BSEs) by neglecting the moir\'e potential and interlayer coupling first.
The IX states for the upper bilayer ($\ell=-1$) or lower bilayer ($\ell=1$) take the form:
\begin{equation}\label{potential}
\begin{split}
\left|\rm{IX}\right\rangle_{\ell,\boldsymbol{Q}}=\sum_{\textit{\textbf{k}}}f_{\ell,\textit{\textbf{Q}}}(\textit{\textbf{k}} )\hat{a}^{\dagger}_{c,\textit{\textbf{k}} +\alpha_c\textit{\textbf{Q}}}\hat{a}_{v,\textit{\textbf{k}}-\alpha_v\textit{\textbf{Q}}}\left|0\right\rangle,
\end{split}
\end{equation}
where $\left|0\right\rangle$ represents the state with valence band filled and conduction band empty. $\hat{a}^{\dagger}_{c,\textit{\textbf{k}} +\alpha_c\textit{\textbf{Q}}} \left(\hat{a}_{v,\textit{\textbf{k}} -\alpha_v\textit{\textbf{Q}} }\right)$ is the operator that creates (annihilates) an electron in the conduction (valence) band, and $\textit{\textbf{Q}}$ ($\textit{\textbf{k}}$) is the electron-hole center-of-mass (relative) momentum. $\alpha_c = m_c/M$ ($\alpha_v =m_v/M$) with $m_c$ ($m_v$) being the effective mass of electron (hole) and $M$ denotes the mass of the exciton.
Under the parabolic band approximation, the electron-hole relative-motion wave function $f_{\ell,\textit{\textbf{Q}}} (\textit{\textbf{k}})$ for the lowest exciton energy $\mathcal{E}_{l}\left(\textit{\textbf{Q}}\right)$, is determined by
\begin{widetext}
	\begin{equation}
 \label{BSE}
\left(E_{c,\textit{\textbf{k}}+\alpha_{c}\textit{\textbf{Q}}}-E_{v,\textit{\textbf{k}}-\alpha_{c}\textit{\textbf{Q}}}\right)f_{l,\textit{\textbf{Q}} }\left(\textit{\textbf{k}}\right)-\sum_{\textit{\textbf{k}}^{\prime}}\left\langle l\textit{\textbf{k}} \textit{\textbf{Q}}\right|V\left|l\textit{\textbf{k}}^{\prime}\textit{\textbf{Q}}\right\rangle f_{l,\textit{\textbf{Q}}}\left(\textit{\textbf{k}}^{\prime}\right)=\mathcal{E}_{l}\left(\textit{\textbf{Q}}\right)f_{l,\textit{\textbf{Q}}}\left(\textit{\textbf{k}}\right),
	\end{equation}
\end{widetext}
where 
$E_{c,\boldsymbol{k}}=\hbar^2\textit{\textbf{k}}^2/2m_c+E_g$, $E_{v,\textit{\textbf{k}}}=-\hbar^2\textit{\textbf{k}}^2/2m_v$ and $E_g$ is the band gap. 
$\left|l\textit{\textbf{k}}\textit{\textbf{Q}}\right\rangle \equiv \hat{a}^{\dagger}_{c,\textit{\textbf{k}}+\alpha_c\textit{\textbf{Q}}}\hat{a}_{v,\textit{\textbf{k}}-\alpha_c\textit{\textbf{Q}}}\left|0\right\rangle$, and
$\left\langle l\textit{\textbf{k}}\textit{\textbf{Q}}\right|V\left|l\textit{\textbf{k}}^{\prime}\textit{\textbf{Q}}\right\rangle = \frac{2\pi e^2}{\mathcal{A}\epsilon\left|\textit{\textbf{k}}-\textit{\textbf{k}}^{\prime}\right|}e^{-\left|\textit{\textbf{k}}-\textit{\textbf{k}}^{\prime}\right|d}$ is the interlayer Coulomb interaction for electron-hole pairs with $\mathcal{A}$ being the system area. 
The wave function is normalized by the condition $\sum_{k}\left|f_{\ell,\textit{\textbf{Q}}}\right|^2=1$.
We choose $m_c = 0.8m_e$ \cite{larentis2018large}, $m_v = 0.43m_e$ \cite{devakul2021magic}, effective dielectric constant $\epsilon=3.8$, and interlayer distance $d=0.67$ nm in the calculations. As the lowest energy exciton state occurs at $\boldsymbol{Q}=0$,
by solving Eq. (\ref{BSE}) for $\boldsymbol{Q}=0$, we find that the exciton binding energy $E_B=E_g-\mathcal{E}_{l} \approx 165 $ meV.

\section{continuum model and topological exciton band}
\label{sec3}
We first construct the continuum model which describes the moir\'e potential formed in this twisted heterotrilayer system. The preservation of both $\hat{C}_3$ and mirror symmetries along with the large energy offset between the conduction and valence bands for middle and outer two layers dictates that the moir\'e potential for the two valence bands from outer two layers is similar to the moir\'e potential of twisted homobilayer TMDs (here is twisted bilayer WSe$_2$). At small twist angles, the three high-symmetry stackings AA, AB and AC can be approximated by angle-aligned heterotrilayers with different stacking configurations, which is the local approximation extensively used to construct a continuum model.
Through first-principles calculations of the band structures for the three commensurate stackings (see Appendix A), we confirm that the modulation of the valence bands at the moir\'e scale can indeed be described by a model similar to that used for twisted homobilayer TMDs. This system differs from the twisted heterotrilayers $\rm{WSe_2/WSe_2/MoSe_2}$ \cite{xie2024long}, where the spatially asymmetric interlayer hybridization between the $\rm{MoSe_2}$ layer and the twisted $\rm{WSe_2/WSe_2}$ bilayer breaks the mirror symmetry between two high-symmetry stackings. As a result, we need to develop a more complicated continuum model with additional parameters, supported by first-principles calculations of band structures for the three high-symmetry stackings of heterotrilayers. As shown below, in addition to the advantage that our model can be described using a similarly simple continuum model as those for twisted homobilayers, the symmetry between the two outer layers also makes the exciton band topological without requiring an external electric field.


\begin{figure}[tbp] \centering
	\includegraphics[width=0.99\linewidth]{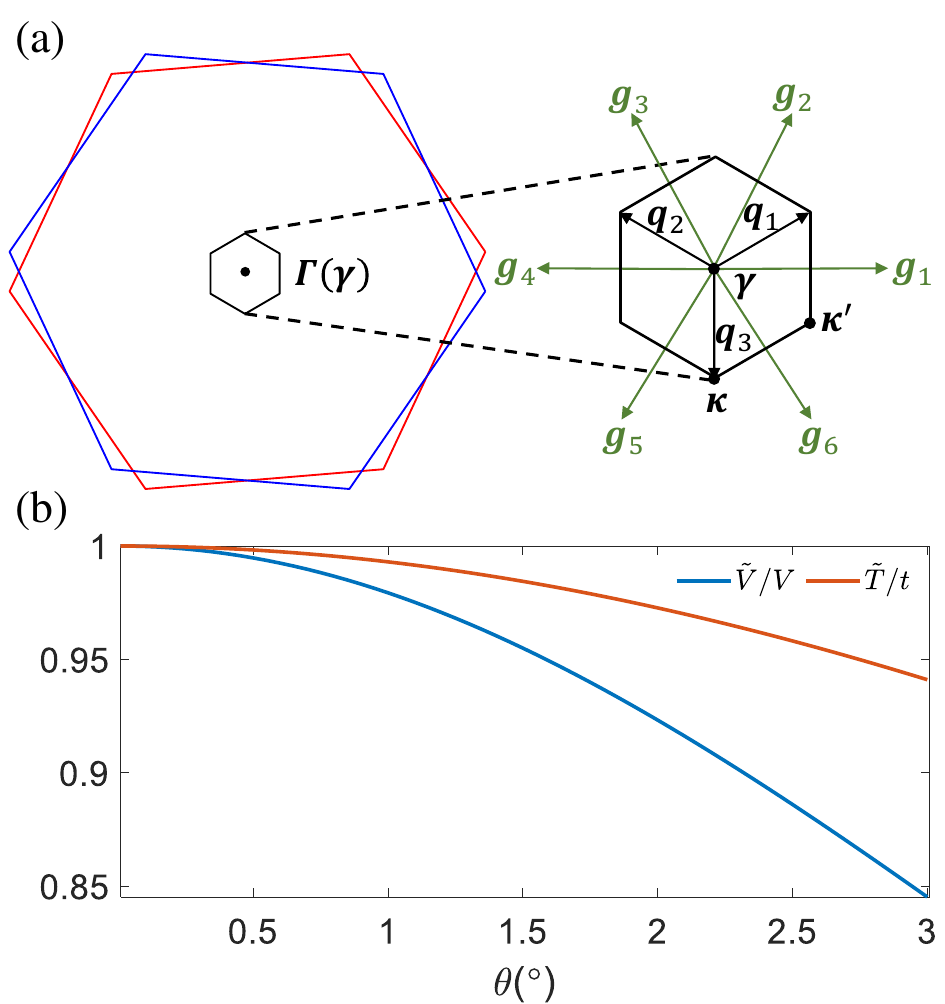}
	\caption{ (a) Brillouin zones of the outer $\rm{WSe_2}$ layers and the corresponding moir\'e Brillouin zone of the $\rm{tWSe_2}$. $\textit{\textbf{g}}_{i}$ (green arrows) with $i=1,...,6$ is the moir\'e reciprocal lattice vector and $\textit{\textbf{q}}_{j}$ (black arrows) with $j=1,2,3$ is the momentum shift between the $K$ points of the outer two layers after rotation. (b) $\widetilde{V}/V$ and $\widetilde{T}/t$ as functions of the twist angle $\theta$.}
	\label{fig2}
\end{figure}

The continuum model which describes the modulation of the $+K$ valley valence bands from the outer $\rm{WSe_2}$ layers at moir\'e scale reads,
\begin{equation}
\begin{aligned}
&\mathcal{H}_{v} =
\begin{pmatrix}
-\frac{\hbar^2 (\boldsymbol{k} - \boldsymbol{\kappa}_+ )^2}{2m_v} + \Delta_{1}(\boldsymbol{r}) & \Delta_T(\boldsymbol{r}) \\
\Delta_T^\dagger(\boldsymbol{r}) & -\frac{\hbar^2 (\boldsymbol{k} - \boldsymbol{\kappa}_- )^2}{2m_v} + \Delta_{-1}(\boldsymbol{r})
\end{pmatrix}, \\
&\Delta_{1,-1}(\boldsymbol{r}) = 2V \sum_{j=1,3,5} \cos(\boldsymbol{g}_j \cdot \boldsymbol{r} \pm \phi), \\
&\Delta_T(\boldsymbol{r}) = t \left( 1 + e^{-i \boldsymbol{g}_2 \cdot \boldsymbol{r}} + e^{-i \boldsymbol{g}_3 \cdot \mathbf{r}} \right).
\end{aligned}
\end{equation}
 Here $\boldsymbol{\kappa}_{\pm}$ are the momenta at the corners of moir\'e Brillouin zone (MBZ) and $\textit{\textbf{g}}_i$ is the moir\'e reciprocal lattice vector obtained by counterclockwise rotation of $\textit{\textbf{g}}_1=(4\pi/\sqrt{3}a_M,0)$ by the angle $(i-1)\pi/3$ [see Fig. \ref{fig2} (a)]. From our first-principles calculations, we obtain the continuum model parameters $\left(t,V,\phi\right) \approx $ (2meV, 7.8meV, $65.7^{\circ}$) for twisted $\rm{WSe_2/MoSe_2/WSe_2}$ (see Appendix A). As expected, since the distance between the outer two layers is larger than the case of twisted homobilayers, the interlayer tunneling between the outer two layers is weaker than the homobilayer case \cite{wu2019topological,devakul2021magic}. The valence band edges splitting at AA stacking is $6 t\approx12$meV, consistent with previous theoretical calculations and experimental observations for the systems with angle alignment \cite{du2024new,yu2023observation,li2023quadrupolar,bai2023evidence,lian2023quadrupolar,xie2023bright,slobodkin2020quantum}.

Denote $\rm{IX}_1$ and $\rm{IX}_{-1}$ as the IX states located at the lower and upper bilayers, corresponding to two pseudospin states, respectively. With the moir\'e continuum model describing the hybridized valence bands, one can calculate the coupling between $\rm{IX}_1$ and $\rm{IX}_{-1}$ brought by valence bands moir\'e potential, and obtain the exciton moir\'e Hamiltonian in momentum space, following the approach detailed in Ref. \cite{xie2024long}.
The exciton moir\'e Hamiltonian in the basis $\left\{\left|\rm{IX}_1\right>,\left|\rm{IX}_{-1}\right>\right\}$, constructed as described in \cite{xie2024long}, is given by
\begin{equation}\label{Hamiltonian}
\begin{split}
&\mathcal{H}_{\textit{\textbf{Q}},\textit{\textbf{Q}}^\prime}=\left(\begin{array}{cc}
\mathcal{E}_{1}\left(\textit{\textbf{Q}}\right)+\mathcal{U}^{1}_{\textit{\textbf{Q}},\textit{\textbf{Q}}^\prime}& \mathcal{T}_{\textit{\textbf{Q}},\textit{\textbf{Q}}^\prime}\\
\left[\mathcal{T}_{\textit{\textbf{Q}},\textit{\textbf{Q}}^\prime}\right]^{\dagger} & \mathcal{E}_{-1}\left(\textit{\textbf{Q}}\right)+\mathcal{U}^{-1}_{\textit{\textbf{Q}},\textit{\textbf{Q}}^\prime}
\end{array}\right),\\
&\mathcal{U}^{l}_{\textit{\textbf{Q}},\textit{\textbf{Q}}^\prime}=-\widetilde{V}_l e^{i s_{l}\phi}\sum^{6}_{i=1}\delta_{\textit{\textbf{Q}}-\textit{\textbf{g}}_i,\textit{\textbf{Q}}'},\\
&\mathcal{T}_{\textit{\textbf{Q}},\textit{\textbf{Q}}^\prime}=\sum^{3} _{n=1}\widetilde{T}_n\delta_{\textit{\textbf{Q}}+\textit{\textbf{q}}_n,\textit{\textbf{Q}}^\prime},
\end{split}
\end{equation}
where $\mathcal{E}_{\ell}\left(\textit{\textbf{Q}}\right)$ is the lowest exciton energy for bilayer $\ell$, and $\mathcal{U}^{l=\pm1}_{\textit{\textbf{Q}},\textit{\textbf{Q}}^\prime}$ is the diagonal moir\'e potential experienced by IX$_1$ and IX$_{-1}$ states, respectively. $\textit{\textbf{q}}_{j}$ is the momentum shift between IX$_1$ and IX$_{-1}$ states, related to each other by a rotation of $2\pi/3$. Neglecting the dependence of $f_{l,\boldsymbol{Q}}$ on $\boldsymbol{Q}$ for simplicity,
$\widetilde{V}_l\approx V\sum_{\textit{\textbf{k}}}f^{*}_{l,0}(\textit{\textbf{k}})f_{l,0}(\textit{\textbf{k}}+\alpha_c\textit{\textbf{g}}_1)$ characterizes the amplitude of $\mathcal{U}^{l}_{\textit{\textbf{Q}},\textit{\textbf{Q}}^\prime}$, $\mathcal{T}_{\textit{\textbf{Q}},\textit{\textbf{Q}}^\prime}$ is the off-diagonal coupling between IX$_1$ and IX$_{-1}$ states brought by interlayer coupling, and
$\widetilde{T}_n\approx t\sum_{\textit{\textbf{k}}}f^{*}_{1,0}(\textit{\textbf{k}})f_{-1,0}(\textit{\textbf{k}}-\alpha_c\textit{\textbf{q}}_n)$ describes the strength of the IX$_1$-IX$_{-1}$ coupling. 
The mirror symmetry between AB and AC stackings ensures that the amplitudes $\widetilde{V}_1$ and $\widetilde{V}_{-1}$ for $\rm{IX}_1$ and $\rm{IX}_{-1}$, respectively, are equal, while their phases are opposite, i.e., $\widetilde{V}_1 = \widetilde{V}_{-1}=\widetilde{V}$ and $s_{l}\phi=-\ell \phi$.
Additionally, the $\hat{C}_3$ symmetry ensures that the hopping strengths satisfy the condition $\widetilde{T}_1 = \widetilde{T}_2 = \widetilde{T}_3$, denoted as $\widetilde{T}$. More details about the derivation of $\mathcal{U}_{\textit{\textbf{Q}},\textit{\textbf{Q}}^\prime}^l$ and $\mathcal{T}_{\textit{\textbf{Q}},\textit{\textbf{Q}}^\prime}$ are available in the supplementary materials of Ref. \cite{xie2024long}. Note that
the exciton moir\'e Hamiltonian discussed above is derived using the continuum approximation, which is only applicable for small twist angles. 

\begin{figure}[tbp] \centering
	\includegraphics[width=0.99\linewidth]{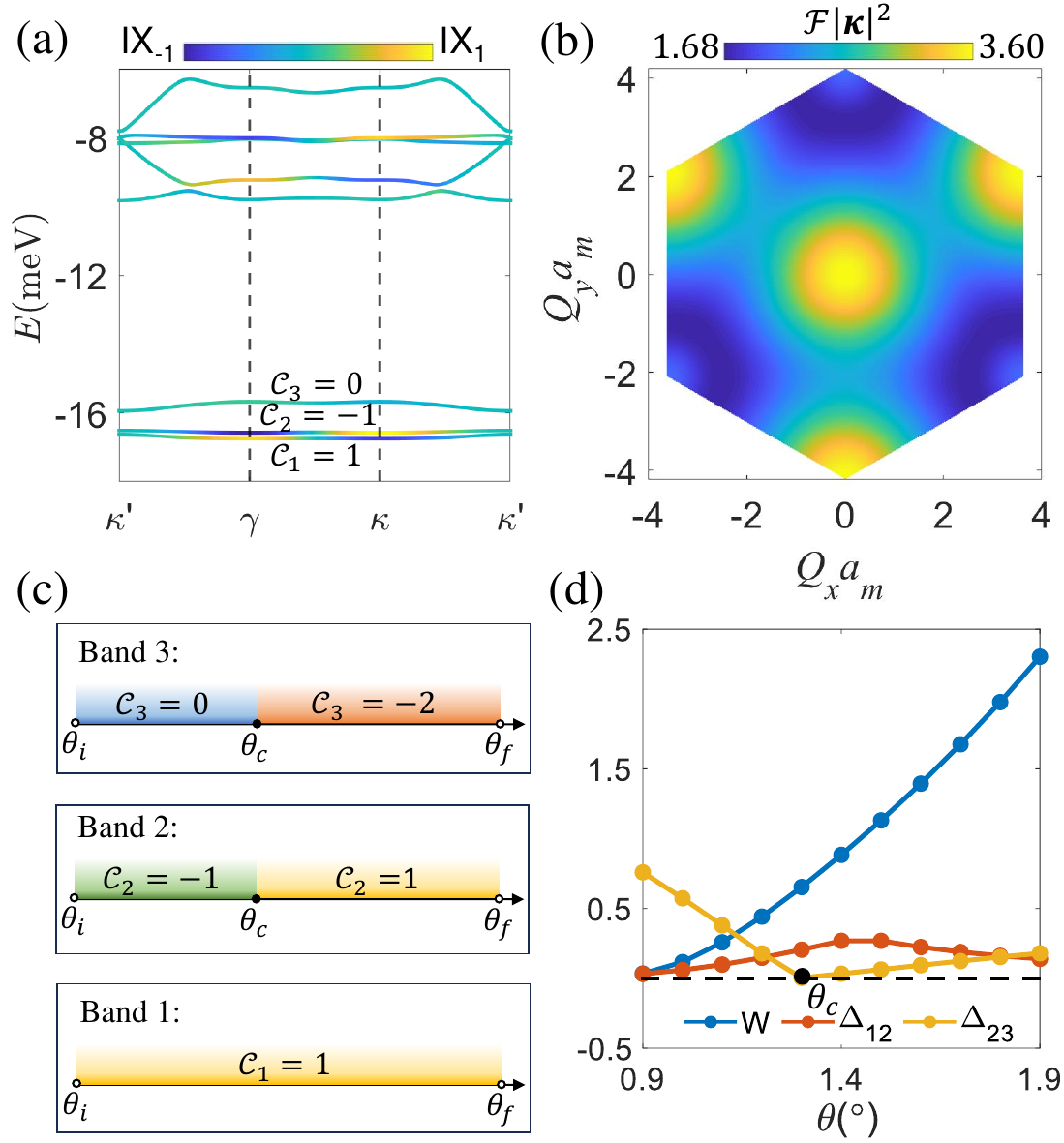}
	\caption{ (a) Exciton moir\'e band structure for $\theta = 1^\circ$ along with the Chern numbers of the lowest three bands. The line color represents the proportion of the $\rm{IX}_1$ and $\rm{IX}_{-1}$ components. 
 (b) Berry curvature $\mathcal{F}\left|\bm{\kappa}\right|^{2}$ of the lowest exciton band in (a). 
 (c) Chern number phase diagram of the lowest three bands. 
 (d) Bandwidth $W$ of the lowest exciton moir\'e band and band gaps $\Delta_{12}$ and $\Delta_{23}$ as functions of $\theta$. The energy unit is meV. $\theta_i=0.9^\circ$, $\theta_f=1.9^\circ$, $\theta_{c}\approx1.3^\circ$.}
	\label{fig3}
\end{figure}

Figure. \ref{fig2}(a) shows the MBZ and the rotated Brillouin zone (RBZ) of the outer two layers. 
We align the center of the MBZ, denoted as $\bm{\gamma}$, with the center of the RBZ, denoted as $\bm{\Gamma}$.
The exciton center-of-mass momenta of $\rm{IX}_1$ and $\rm{IX}_{-1}$ are measured from $\bm{\gamma}$ and $\bm{\kappa}$ points, respectively.
Figure. \ref{fig2}(b) shows $\widetilde{V}$ and $\widetilde{T}$ as functions of the twist angle $\theta$.
The parameters $\widetilde{V}$ and $\widetilde{T}$ decrease monotonically with increasing twist angle, because they depend on the overlap of exciton wave functions with momenta differing by $\boldsymbol{g}_j$ and $\boldsymbol{q}_j$, and this overlap is suppressed for large momenta transfer corresponding to large twist angles.

Based on the continuum model parameters, we numerically diagonalize the exciton moir\'e Hamiltonian and plot the exciton moir\'e bands $E_n(\textit{\textbf{k}})$ in MBZ for a typical angle $\theta=1^\circ$, as shown in Fig. \ref{fig3}(a).
Here $n$ is the band index, with the lowest energy band labeled as $n=1$.
The moir\'e band structure consists of three lowest bands that are isolated from the higher-energy bands, with the two lowest bands, $n=1$ and $n=2$, being closer in energy to each other than to the third band, $n=3$. The bandwidth of the lowest band $W$, and the energy gaps $\Delta_{nn'}$ between the $n$th band maximum and $n'$th band minimum, are shown in Fig. \ref{fig3}(d).
The weak hybridization between $\rm{IX}_1$ and $\rm{IX}_{-1}$ results in a small energy gap between the $n=1$ and $n=2$ bands.
In the lowest band, the wave functions of $\rm{IX}_1$ and $\rm{IX}_{-1}$ states are localized at the $\bm{\gamma}$ and $\bm{\kappa}$ points, respectively, while this behavior is reversed in band $n=2$.
The topology of exciton moir\'e band can be characterized by the Berry curvature $\mathcal{F}_n$ and Chern number $\mathcal{C}_{n}$ \cite{xiao2010berry}:
\begin{equation}\label{BerryandChern}
\begin{split}
&\mathcal{F}_n(\textit{\textbf{Q}})=\hat{z}\nabla_{\textit{\textbf{Q}}}\times[i\left<\chi_n(\textit{\textbf{Q}})\right|\nabla_{\textit{\textbf{Q}}}\left|\chi_n(\textit{\textbf{Q}})\right>],\\
&\mathcal{C}_n= \frac{1}{2\pi}\int_{\rm{MBZ}} {d^2\textit{\textbf{Q}}}\mathcal{F}_n(\textit{\textbf{Q}}),
\end{split}
\end{equation}
where $\chi_{n}(\textit{\textbf{Q}})$ is the eigenstate of Eq. (\ref{Hamiltonian}) for the $n$th band.
Figures \ref{fig3}(a)-(c) show the topological properties of the IX bands.
The lowest two moir\'e bands have nonzero Chern numbers, while the third band can be either topologically trivial or nontrivial, depending on the twist angle.
The Berry curvature $\mathcal{F}_1$ has hot spots around the high-symmetry points $\bm{\gamma}$, $\bm{\kappa}$ and $\bm{\kappa^\prime}$, as a result of the hybridization between $\rm{IX}_1$ and $\rm{IX}_{-1}$.
Unlike the asymmetric configuration \cite{xie2024long}, the lowest two bands remain topologically nontrivial even in the absence of external electric field.
Applying an electric field creates an energy imbalance between $\rm{IX}_1$ and $\rm{IX}_{-1}$, which suppresses the formation of topological bands. Due to time-reversal symmetry, the moir\'e exciton bands for $-K$ valley excitons have the same dispersion as those for $+K$ valley excitons but with opposite Chern numbers.

According to the values of $\mathcal{C}_n$ for the lowest three moir\'e bands, we construct the topological phase diagram shown in Fig. $\ref{fig3}$(c), which can be divided into two regions: phase I with $(\mathcal{C}_1,\mathcal{C}_2,\mathcal{C}_3)=(1,-1,0)$, and phase II with $(\mathcal{C}_1,\mathcal{C}_2,\mathcal{C}_3)=(1,1,-2)$. 
The topological phase trasition between phase I and phase II occurs at a critical angle, i.e., $\theta_{c}\approx1.3^\circ$.
We focus on the angle range $\theta_i < \theta < \theta_f$, where $\theta_i = 0.9^\circ$ and $\theta_f = 1.9^\circ$. Within this range, the lowest band carries a nonzero Chern number. 
As illustrated in Fig. \ref{fig3}(d), for small twist angle $\theta<\theta_{c}$, the lowest energy bandwidth $W$ and band gap $\Delta_{12}$ increase with $\theta$, while $\Delta_{23}$ decreases monotonically.
When $\theta=\theta_{c}$, the topological phase transition from phase I to phase II occurs, accompanied by the closing of the band gap $\Delta_{23}$ at $\boldsymbol{\kappa}^\prime$ point.
As $\theta$ increases further, $\Delta_{23}$ gradually increases while $\Delta_{12}$ gradually decreases.
In all these phases, the total Chern number of the lowest three moir\'e bands is zero, suggesting that a three-orbital TB model can capture the essence of the lowest three bands. 

Since the system has $\hat{C}_3$ rotation symmetry, the Bloch states $\ket{\Psi_{n,\boldsymbol{Q}}}$ at the three high-symmetry points $\boldsymbol{\gamma}$, $\boldsymbol{\kappa}$, and $\boldsymbol{\kappa}^\prime$ have definite the angular momenta denoted as $L_{n,\boldsymbol{Q}}$ of (see Appendix B for details). It is known that in a system with $\hat{C}_3$ rotation symmetry, the Chern numbers $\mathcal{C}_n$ and angular momenta $L_{n,\boldsymbol{Q}}$ satisfy the condition $[\mathcal{C}_n-(L_{n,\boldsymbol{\gamma}}+L_{n,\boldsymbol{\kappa}}+L_{n,\boldsymbol{\kappa}'})]\: \text{mod}\: 3=0$ \cite{fang2012bulk}.
Using the values of $L_{n,\boldsymbol{\gamma}}$, $L_{n,\boldsymbol{\kappa}}$ and $L_{n,\boldsymbol{\kappa}'}$ listed in Table \ref{table-2}, it can be verified that the above relationship is satisfied in our calculations.

\section{Three-orbital tight-binding model}
\label{sec4}
In order to understand the topological properties of the lowest exciton bands and also facilitate the following many-body exciton phase calculations, we now construct a TB model with the Wannier functions.
Within the angle range $\theta_i\textless\theta\textless\theta_f$, the total Chern number for the lowest three bands satisfies $\sum_{n=1,2,3}\mathcal{C}_n=0$, allowing for the construction of a unified three-orbital TB model to describe phase I and phase II.
The Wannier functions for the lowest three moir\'e bands at $\boldsymbol{R}=0$ can be expressed as
\begin{equation}\label{wannier}
    \begin{split}
    \ket{W_{\alpha}}=\frac{1}{\sqrt{N}}\sum_{n=1,2,3}\sum_{\boldsymbol{Q}}V^{n}_{\alpha,\boldsymbol{Q}}\ket{\Psi_{n,\boldsymbol{Q}}},
    \end{split}
\end{equation}
where $\alpha$ labels the three Wannier orbitals ($\alpha=\{1,2,3\}$ is also interchangeably labeled as $\{A,B,O\}$), $N$ is the total number of $\boldsymbol{Q}$ in the MBZ, $n$ is the exciton band index, $\ket{\Psi_{n,\boldsymbol{Q}}}$ is the Bloch state of $n$th band, and $V_{\boldsymbol{Q}}$ is a $3\times3$ unitary matrix introduced to disentangle the exciton layer hybridization.
Denote the wave function $\sum_{n=1,2,3}V^{n}_{\alpha,\boldsymbol{Q}}\ket{\Psi_{n,\boldsymbol{Q}}}$ as $\ket{\Phi_{\alpha,\boldsymbol{Q}}}$ for convenience in the following discussions.
For $\alpha=A$, $V^{n}_{\alpha,\boldsymbol{Q}}$ is obtained by requiring that $\ket{\Phi_{\alpha,\boldsymbol{Q}}}$ is maximally polarized to the $\rm{IX}_{1}$ component, while for $\alpha=B$, it is maximally polarized to the $\rm{IX}_{-1}$ component.
This optimization procedure is equivalent to solving the eigenstates of the polarization operator $\hat{\sigma}_z$ matrix under the Bloch state basis $\ket{\Psi_{n,\boldsymbol{Q}}}$ \cite{qiu2023interaction}. 

The overall phase of $\ket{\Phi_{\alpha,\boldsymbol{Q}}}$ must be specified through additional requirements.
We choose the phase of $\ket{\Phi_{A,\boldsymbol{Q}}}$ ($ \ket{\Phi_{B,\boldsymbol{Q}}}$) to ensure that its $\rm{IX}_1$ ($\rm{IX}_{-1}$) component $\ket{\Phi^{1}_{A,\boldsymbol{Q}}}$ ($\ket{\Phi^{-1}_{B,\boldsymbol{Q}}}$) is real and positive at the position $\boldsymbol{r}_{A}$ ($\boldsymbol{r}_{B}$).
The phase of $\ket{\Phi_{O,\boldsymbol{Q}}}$ is determined by first fixing the gauge at $\boldsymbol{Q}=\boldsymbol{\gamma}$, such that  $\ket{\Phi_{O,\boldsymbol{\gamma}}}\textgreater0$ and
$\langle\Phi_{O,\boldsymbol{\gamma}}|\Phi_{O,\boldsymbol{Q}}\rangle>0$ at $\boldsymbol{r}_{O}$ for any $\boldsymbol{Q}$. Here
the positions $\boldsymbol{r}_{A}=a_M(1/\sqrt{3},0)$, $\boldsymbol{r}_{B}=a_M(2/\sqrt{3},0)$, and $\boldsymbol{r}_{O}=a_M(0,0)$, represent the locations of the high-symmetry locals with AB, AC, and AA stackings within the moir\'e unit cell, respectively.
The amplitude of the Wannier state $W_{\alpha}\left(\boldsymbol{r}\right)=\left[W^{1}_{\alpha}\left(\boldsymbol{r}\right),W^{-1}_{\alpha}\left(\boldsymbol{r}\right)\right]^{T}$, centered at $\boldsymbol{r}=\boldsymbol{r}_{\alpha}$, is shown in Fig. \ref{fig4}.
The Wannier states at the A and B sublattices exhibit dominant contributions in the $W^{1}_{\alpha}\left(\boldsymbol{r}\right)$ and $W^{-1}_{\alpha}\left(\boldsymbol{r}\right)$  components, respectively, signifying two excitonic modes with opposite dipole moments. In contrast, the O sublattice Wannier state demonstrates a symmetric distribution across both components, indicative of a quadrupole exciton.


\begin{figure}[tbp] \centering
\includegraphics[width=0.99\linewidth]{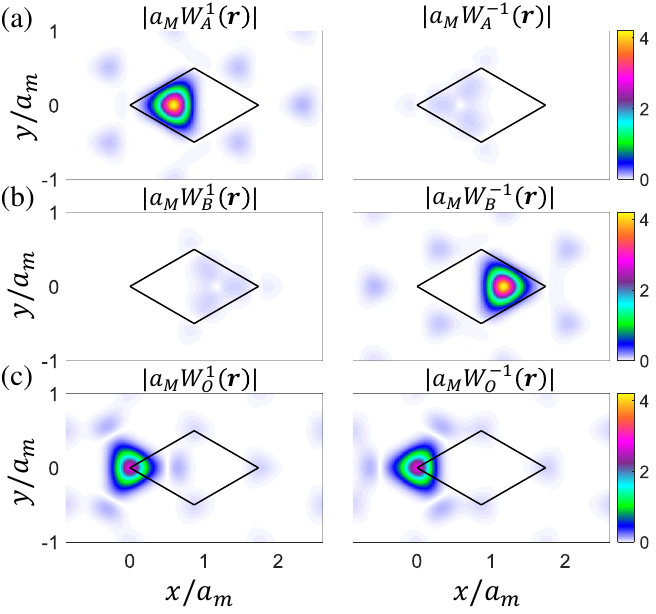}
\caption{ (a)-(c) Spatial distribution of the amplitudes of three Wannier orbitals constructed for $\theta=1^\circ$.
$W_{\alpha}^{\ell}$ with $\alpha=\{A, B, O\}$ and $\ell=\{1,-1\}$, represents the $\rm{IX}_\ell$ component of the Wannier state for the $\alpha$ orbital. 
The black rhombus indicates the moir\'e unit cell at $\boldsymbol{R}=0$.}
\label{fig4}
\end{figure}


Keeping the nearest-neighbor (NN) and next-nearest-neighbor (NNN) hopping terms, we now build a single-particle TB model restricted to lowest three exciton bands, described by the Hamiltonian
\begin{equation} \label{TB}
  \begin{split}
    \hat{H}_{0}=&-\sum_{\alpha}\sum_{i}t^{(0)}_{i_\alpha,i_\alpha}\hat{b}_{i_\alpha}^{\dagger}\hat{b}_{i_\alpha}-\sum_{\alpha\atop\left(\alpha\neq\alpha^\prime\right)}\sum_{\left\langle i,j\right\rangle }t^{(1)}_{i_\alpha,j_{\alpha^\prime}}\hat{b}_{i_\alpha}^{\dagger}\hat{b}_{j_{\alpha^\prime}}\\
    &-\sum_{\alpha}\sum_{\left\langle\left\langle i,j\right\rangle\right\rangle}t^{(2)}_{i_\alpha,j_\alpha}\hat{b}_{i_\alpha}^{\dagger}\hat{b}_{j_\alpha},
  \end{split}
\end{equation}
where $\hat{b}_{i_\alpha}$ ($\hat{b}_{i_\alpha}^{\dagger}$) is the annihilation (creation) operator for IXs at lattice $\boldsymbol{R}_i$ associated with the Wannier orbital $\alpha$, $\left\langle...\right\rangle$ denotes the NN lattice sites, and $\left\langle\left\langle...\right\rangle\right\rangle$ labels the NNN lattice sites. 

The first term in $\hat{H}_{0}$ consists of on-site energy shifts of three orbitals of IXs, with the purely real $t^{(0)}_{i_\alpha,i_\alpha}$. The remaining two terms in $\hat{H}_{0}$ describe NN inter-orbital and NNN intra-orbital hopping processes, characterized by the amplitudes $t^{(1)}_{i_\alpha,j_\alpha^\prime}$ and $t^{(2)}_{i_\alpha,j_\alpha}$, respectively, which are generally complex-valued.
The hopping parameter $t_{i_\alpha,j_\beta}$ is obtained by calculating the exciton Hamiltonian matrix element within the Wannier basis $\left\{W_{A},W_{B},W_{O}\right\}$, with the expression
\begin{equation} \label{hopping}
  \begin{split}
t_{i_\alpha,j_\beta}&\equiv-\bra{W_{\alpha,\boldsymbol{R}_{m}}}\mathcal{H}\ket{W_{\beta,\boldsymbol{0}}}\\
&=-\frac{1}{N}\sum_{\boldsymbol{Q}}e^{i\boldsymbol{Q}\cdot\boldsymbol{R}_{m}}\sum_{n}\left[V_{\alpha,\boldsymbol{Q}}^{n}\right]^{*}E_n\left(\boldsymbol{Q}\right)V_{\beta,\boldsymbol{Q}}^{n},
  \end{split}
\end{equation}
which describes the hopping between a site at $\boldsymbol{R}_{m}$ belonging to $\alpha$ orbital and a site at $\boldsymbol{R}=\boldsymbol{0}$ belonging to $\beta$ orbital, and $E_n(\boldsymbol{Q})$ is the eigenvalue of Eq. (\ref{Hamiltonian}). 
The hopping processes and their associated phase factors for the three orbitals, as well as the honeycomb lattice are illustrated in Fig. \ref{fig5}(a).
We show the absolute value of $t_{i_\alpha,j_\beta}$ as a function of twist angle $\theta$ in Fig. \ref{fig5}(c).
For small $\theta$, IXs tends to be localized at the corresponding lattice sites and the amplitude of inter-orbital hopping $t_{i_\alpha,j_\beta}$ is suppressed. 

By plugging Eq. (\ref{hopping}) into Eq. (\ref{TB}) and diagonalizing the non-interacting Hamiltonian, we obtain the dispersion of the three-orbital TB model. 
As illustrated in Fig. \ref{fig5}(b), the dispersion of TB model agrees well with the low-energy band structure of the continuum model, indicating that the three-orbital TB model can faithfully reproduce the single-particle properties of the continuum model.
\begin{figure}[tbp] \centering
	\includegraphics[width=0.99\linewidth]{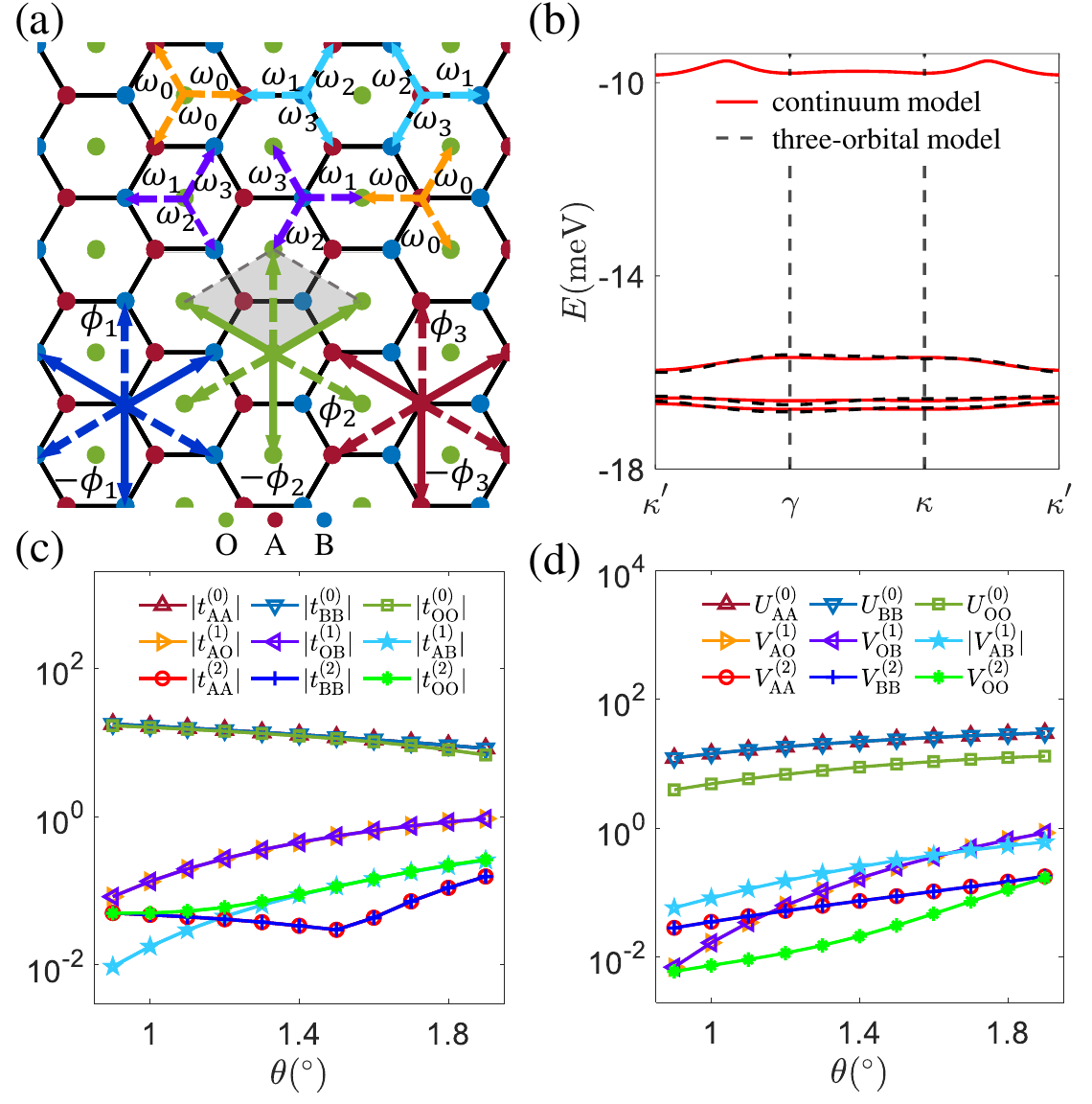}
	\caption{(a) Schematic illustration of the three-orbital TB model with NN and NNN hoppings. The shaded rhombus represents the moir\'e unit cell at $\boldsymbol{R}=0$. $\omega_{i}$ is the phases of NN  $(t^{(1)}_{\alpha,\beta})$ and NNN $(t^{(2)}_{\alpha,\alpha})$  hopping parameter. Here $\omega_{0}=0$, $\omega_{1}=0$, $\omega_{2}=e^{i2\pi/3}$ and $\omega_{3}=e^{-i2\pi/3}$. 
    $\phi_{i}$ is the spatially dependent phase of NNN $(t^{(2)}_{\alpha,\alpha})$ hopping parameter. The dashed and solid lines connecting two identical orbitals exhibit identical NNN hopping phases.
    In the limit of zero interlayer hybridization, $\phi_{1}=2\pi/3$, $\phi_{2}=-2\pi/3$ and $\phi_{3}=0$.
 (b) Band dispersion from the three-orbital TB model (black dashed line) and the continuum model (red solid line) for $\theta=1^\circ$. 
 (c) The absolute values of on-site energy $(|t^{0}_{\alpha,\alpha}|)$, NN $(|t^{1}_{\alpha,\beta}|)$ and NNN $(|t^{2}_{\alpha,\alpha}|)$ hopping parameters, as functions of $\theta$.
 (d) The values of on-site $(U_{\alpha})$, NN $(V_{\alpha,\beta})$ and NNN $(V_{\alpha,\alpha})$ interactions as functions of $\theta$. 
In particular, $V_{A,B}=-|V_{A,B}|$ corresponds to NN attractive interaction. The energy unit in (c)-(d) is meV.}
	\label{fig5}
\end{figure}

\section{many-body phase diagram of interlayer excitons}
\label{sec5}
Based on the three-orbital TB model, we now turn to study the many-body properties of the excitons in this moir\'e lattice. For small twist angles, the moir\'e supercell size is much larger than the exciton Bohr radius, and thus we can treat the IXs as composite bosonic particles interacting via the dipole-dipole interaction \cite{wu2018theoryPRB,gotting2022moire}. In particular, according to the properties of Wannier orbitals, excitons in Wannier orbitals $W_A(\boldsymbol{r})$ and $W_B(\boldsymbol{r})$ are dipolar excitons with opposite dipole moments, while excitons in $W_O(\boldsymbol{r})$ are quadrupolar excitons. The exciton density can be conveniently tuned in experiments by varying the laser excitation power. Additionally, the polarization of the laser can be used to selectively control exciton densities in different valleys. In the following, we focus only on the excitation of excitons in the $+K$ valley.
Based on the three-orbital TB model, the interacting IXs can be described by an extended moir\'e Bose-Hubbard (EMBH) model which takes the form
\begin{equation} \label{BHmodel}
  \begin{split}
    \hat{H}=& \hat{H}_{0}+\sum_{\alpha,\alpha^\prime\atop \alpha\neq\alpha^\prime}\sum_{\left\langle i,j\right\rangle }\frac{V_{\alpha,{\alpha^\prime}}}{2}n_{i_{\alpha}}n_{j_{\alpha^\prime}}+\sum_{\alpha}\sum_{\left\langle \left\langle i,j\right\rangle \right\rangle}\frac{V_{\alpha,\alpha}}{2}n_{i_{\alpha}}n_{j_{\alpha}} \\
    & +\sum_{\alpha}\sum_{i}\frac{U_\alpha}{2}n_{i_{\alpha}}\left(n_{i_{\alpha}}-1\right),
  \end{split}
\end{equation}
where $V_{\alpha,\alpha^\prime}$ is the NN interaction between orbitals $\alpha$ and $\alpha^\prime$, $V_{\alpha,\alpha}$ is the interaction between the NNN sites of the same orbital $\alpha$, and $U_{\alpha}$ is the on-site interaction of the orbital $\alpha$. Since excitons in Wannier orbitals $W_{A}(\boldsymbol{r})$, $W_{B}(\boldsymbol{r})$ and $W_{O}(\boldsymbol{r})$ correspond to dipolar excitons with opposite dipole moments and quadrupolar excitons respectively, the interaction strengths between IXs are related with the following interaction potentials \cite{slobodkin2020quantum},
\begin{align}
I_{p,p}(\boldsymbol{r}) &= \mathcal{I} \left( \frac{2}{r} - \frac{2}{\sqrt{r^2 + d^2}} \right), \\
I_{p,-p}(\boldsymbol{r}) &= \mathcal{I} \left( \frac{1}{r} + \frac{1}{\sqrt{r^2 + (2d)^2}} - \frac{2}{\sqrt{r^2 + d^2}} \right), \\
I_{q,q}(\boldsymbol{r}) &= \mathcal{I} \left( \frac{3}{2r} + \frac{1}{2\sqrt{r^2 + (2d)^2}} - \frac{2}{\sqrt{r^2 + d^2}} \right), \\
I_{q,p}(\boldsymbol{r}) &= \mathcal{I} \left( \frac{3}{2r} + \frac{1}{2\sqrt{r^2 + (2d)^2}} - \frac{2}{\sqrt{r^2 + d^2}} \right),
\end{align}
where $d$ is the layer distance and the interaction constant $\mathcal{I}={e^2}/{4\pi\epsilon}$. The four interaction potentials above describe the interaction between dipolar excitons with parallel dipole moments, the interaction between dipolar excitons with anti-parallel dipole moments, the interaction between quadrupolar excitons, and the interaction between quadrupolar excitons and dipolar excitons, respectively. 

With these interaction potentials, the NN and NNN interaction strengths in Eq. (\ref{BHmodel}) can be expressed as
\begin{align}\label{interaction}
V_{A,A} &=  \iint d\boldsymbol{r}d\boldsymbol{r}^{\prime}\left|W_{i,A}(\boldsymbol{r})W_{j,A}(\boldsymbol{r}^{\prime})\right|^2 I_{p,p}(\boldsymbol{r}-\boldsymbol{r}'), \\
V_{A,B} &=  \iint d\boldsymbol{r}d\boldsymbol{r}^{\prime}\left|W_{i,A}(\boldsymbol{r})W_{j,B}(\boldsymbol{r}^{\prime})\right|^2 I_{p,-p}(\boldsymbol{r}-\boldsymbol{r}'), \\
V_{O,O} &=  \iint d\boldsymbol{r}d\boldsymbol{r}^{\prime}\left|W_{i,O}(\boldsymbol{r})W_{j,O}(\boldsymbol{r}^{\prime})\right|^2 I_{q,q}(\boldsymbol{r}-\boldsymbol{r}'), \\
V_{O,A} &=  \iint d\boldsymbol{r}d\boldsymbol{r}^{\prime}\left|W_{i,O}(\boldsymbol{r})W_{j,A}(\boldsymbol{r}^{\prime})\right|^2 I_{q,p}(\boldsymbol{r}-\boldsymbol{r}').
\end{align}
Other terms can be obtained according to symmetry, i.e., $V_{B,B}=V_{A,A}$, $V_{B,A}=V_{A,B}$, $V_{B,O}=V_{O,B}=V_{A,O}=V_{O,A}$. Similarly, the on-site interaction strengths are
\begin{align}
U_{A} &=  \iint d\boldsymbol{r}d\boldsymbol{r}^{\prime}\left|W_{i,A}(\boldsymbol{r})W_{i,A}(\boldsymbol{r}^{\prime})\right|^2 I_{p,p}(\boldsymbol{r}-\boldsymbol{r}'), \\
U_{O} &=  \iint d\boldsymbol{r}d\boldsymbol{r}^{\prime}\left|W_{i,O}(\boldsymbol{r})W_{i,O}(\boldsymbol{r}^{\prime})\right|^2 I_{q,q}(\boldsymbol{r}-\boldsymbol{r}'), 
\end{align}
and $U_{B}=U_{A}$. Figure \ref{fig5}(d) shows the dependence of various interaction strengths on the twist angle, all of which exhibit a monotonic increase. By comparing Figs. \ref{fig5}(c)-(d) and Fig. \ref{fig3}(d), it becomes evident that several exciton-exciton interaction strengths are significantly stronger than the hopping strengths which approximately represent the bandwidth, and the band gaps between the second and third bands. As a result, the two-band approximation is unsuitable, requiring the use of a three-orbital TB model instead of a two-orbital TB model to accurately describe the many-body phases.

\begin{figure*}[tbp] \centering
\includegraphics[width=0.98\linewidth]{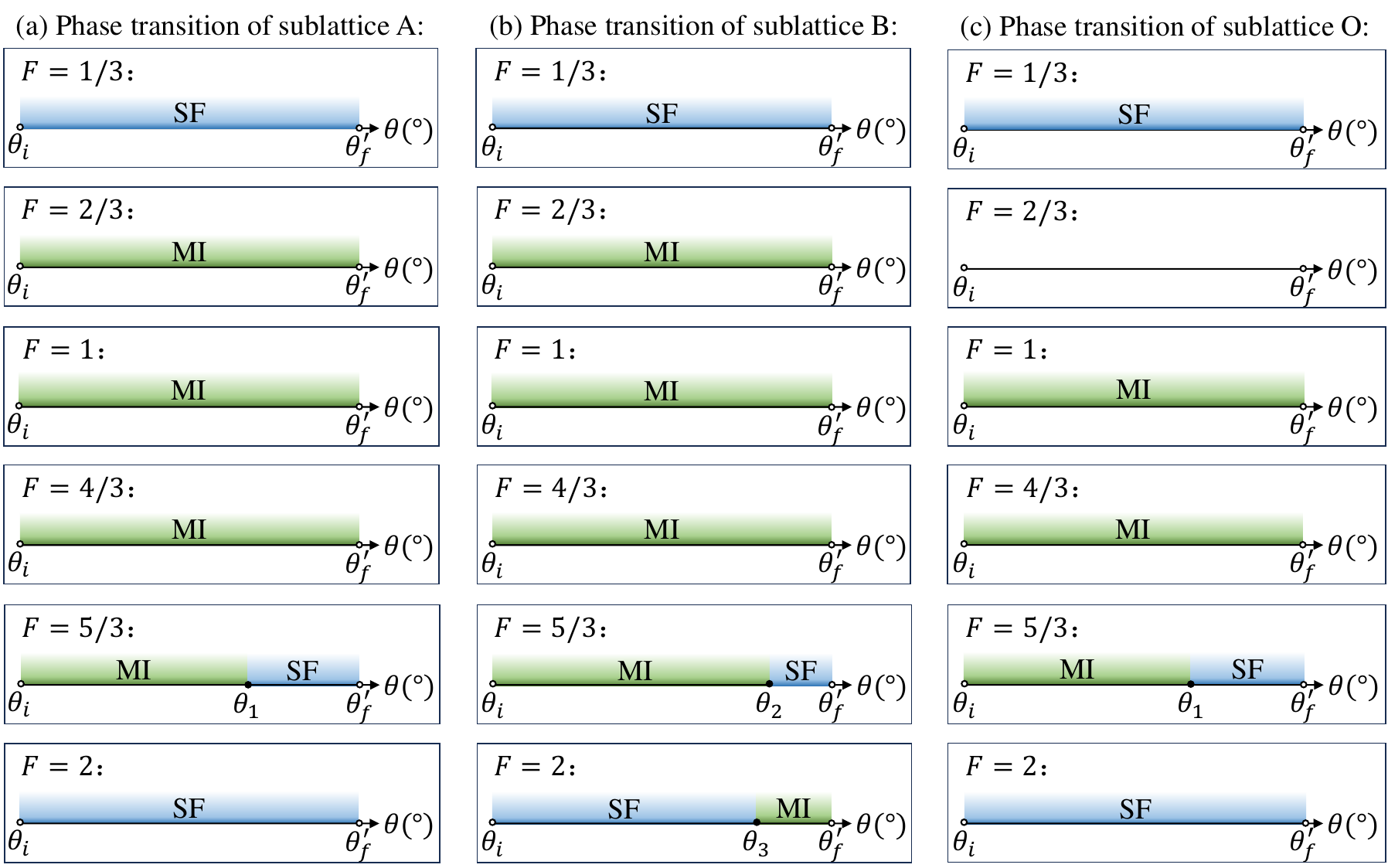}
\caption{(a)-(c) illustrates the many-body phase diagrams as functions of the filling factor $F$ and twist angle $\theta$ at sublattices A, B, and O, respectively.
$\theta_{i}=0.9^\circ$, $\theta_{f^\prime}=1.9^\circ$, $\theta_1=1.4^\circ$, $\theta_2=1.7^\circ$, $\theta_3=1.6^\circ$. }
\label{fig6}
\end{figure*}

Within the framework of mean-field theory, we investigate the ground state of Eq. (\ref{BHmodel}) via the bosonic Gutzwiller variational approach (see Appendix C for more details) \cite{jaksch1998cold,krauth1992gutzwiller,lanata2012efficient,zeng2024}, which can describe the phase transition between an exciton superfluid \cite{deng2022moire,chen2023searching} and Mott insulator \cite{gotting2022moire}.
Assuming that the ground state does not have charge-density-wave order, meaning the exciton wave functions follow the periodicity of the moir\'e superlattice, we numerically compute the ground state of EMBH as a function of twist angle and exciton density (see Fig. \ref{fig6}).

To ensure that the three-band TB model remains applicable, the angle range is further restricted to $0.9^\circ \sim 1.9^\circ$, represented as $\theta_i\sim\theta_f^{\prime}$. Beyond $\theta_f^\prime$, the band gap $\Delta_{34}$ is smaller than the strongest exciton-exciton interaction strength, invalidating the usage of three-band approximation.
The filling factor in a unit cell is defined as $F=N_{t}/N_{s}$, where $N_{t}$ is the number of excitons and $N_{s}=3$ is the number of sublattices within the unit cell. For each sublattice $\alpha$, one can define the corresponding superfluid order parameter $\langle\hat{b}_{\alpha}\rangle$. In the numerical calculations, a superfluid phase in sublattice $\alpha$ is defined as $\langle\hat{b}_{\alpha}\rangle \geq 10^{-2}$. For $\langle\hat{n}_{\alpha}\rangle$ being an integer and $\langle\hat{b}_{\alpha}\rangle<10^{-2}$, the phase is defined as a Mott insulator in sublattice $\alpha$. For integer filling at sublattice $\alpha$, the ground state can be a Mott insulator when the on-site interaction is much stronger compared with the hopping amplitude. This happens at certain fillings, i.e., $F=N/3$ with $N$ being an integer, and for small twist angles. The phase in sublattice $\alpha$ is always a superfluid if its filling is not an integer. Note that superfluid and Mott insulator phases can coexist within a unit cell for different sublattices. 

The calculated ground-state phase diagrams for all sublattices at various fillings and twist angles are shown in Fig. \ref{fig6}. 
At fractional filling of $F=1/3$, where one exciton is shared among the three sublattices, all sublattices exhibit superfluid order due to the dominance of NN hopping over on-site Coulomb interactions. 
At a filling of $F=2/3$, the ground state undergoes a transition to a Mott insulator phase localized on the A and B sublattices, driven by the dominant on-site Coulomb interactions over NN hoppings on these sublattices.  Meanwhile, the higher on-site interaction energy at the O sublattice suppresses excitons occupation, resulting in a sublattice-selective Mott state with unoccupied O sublattices.
For even higher filling, excitons begin to occupy the O sublattices. At integer filling $F=1$, all sublattices enter the Mott insulator phase over the entire range of twist angles considered, with an occupation of $\langle\hat{n}_{\alpha}\rangle=1$ for $\alpha=A,B,O$.
For the filling of $F=4/3$, the ground state remains a Mott insulator but with different sublattice occupations: $\langle\hat{n}_{A}\rangle=\langle\hat{n}_{B}\rangle= 1$ and $\langle\hat{n}_{O}\rangle=2$.
At higher fillings $F=5/3$ and $F=2$, the increased exciton density favors the superfluid phase compared to the case of $F=4/3$, particularly at large twist angles. The superfluid phase region expands, and various superfluid to Mott insulator phase transitions occur at different twist angles for the three sublattices. This aligns with the conventional understanding of the Bose-Hubbard model, where higher boson filling leads to stronger boson number fluctuations, favoring the superfluid phase.
Notably, excitons occupying A and B sublattices have dipoles pointing in opposite directions. 
As a result, the ground state of the system can be characterized as a staggered dipolar superfluid or Mott insulator, depending on the order parameters at the A and B sublattices \cite{slobodkin2020quantum}.

\section{discussion and conclusion}
\label{sec6}
For the Mott insulator phase identified here, where excitons occupy topological bands, additional calculations are required to uncover its potential topological properties. In addition, as a key difference between excitons in the superfluid and Mott insulator phases is their coherence behavior, these two phases predicted in this work can be experimentally distinguished by measuring the photon coherence emitted by the excitons \cite{cutshall2025imaging}.

The exciton topological band is not flat in terms of the band gap to bandwidth ratio. This is primarily due to the suppression of tunneling between the two outer layers, as the weak coupling between the valence bands is unfavorable for the formation of flat bands. Further extention to other heterostructures which preserve the symmetric stackings could lead to the formation of a topological flat band for excitons, potentially enabling the realization of bosonic topological phases in exciton systems. This advancement could offer a new platform for studying these intriguing phases and significantly expand the range of systems available for investigating bosonic topological orders.

In summary, we have proposed twisted symmetric TMD heterotrilayers as a platform for realizing exciton topological bands without the need for an external electric field. The heterostructures are based on similar systems which have been experimentally realized recently, by introducing the twist angle. We discover that within a specific range of twist angles, the lowest three exciton bands acquire topological properties with nonzero Chern numbers. We further construct a three-orbital TB model, which accurately captures the exciton band structures. Building on this model, we study the many-body exciton phases by filling the moir\'e superlattices with dipolar and quadrupolar excitons. By varying the exciton filling and twist angle, we identify sublattice-dependent superfluid and Mott insulator phases as well as their phase transitions. Our study demonstrates that twisted symmetric TMD heterotrilayers provide a realistic platform for exploring exciton topological bands and novel many-body exciton phases, paving the way toward the realization of many-body topological exciton phases in twisted TMD systems.

\begin{acknowledgements}
This work is supported by the National Key Research and Development Program of China (Grant No. 2022YFA1405304), the National Natural Science Foundation of China (Grant Nos. 12004118, 12404181 and 12274145), the Guangdong Provincial Quantum Science Strategic Initiative (Grant No. GDZX2401002), Guangdong Basic and Applied Basic Research Foundation (Grant No. 2023A1515010672) and Guangdong Provincial University Science and Technology Program (Grant No. 2023KTSCX029).
\end{acknowledgements}

\appendix
\section{FIRST-PRINCIPLES CALCULATIONS}
\label{app:First-principles}
The parameters in the valence band continuum model are obtained by performing first-principles calculations for three commensurate stackings AA, AB and AC. First-principles calculations were conducted using density functional theory (DFT) \cite{hohenberg1964inhomogeneous,kohn1965self}, implemented in the Vienna $\emph{ab initio}$ simulation package (VASP) \cite{kresse1996efficient,kresse1996efficiency}. The projector-augmented wave (PAW) \cite{Blochl1994Projector} method was employed to describe electron core interactions, with a cutoff energy of 450 eV. The Perdew-Burke-Ernzerhof (PBE) formalism within the generalized gradient approximation (GGA) was used to treat the exchange correlation \cite{Perdew1996Generalized,Perdew1997Generalized}. The Brillouin zone was sampled using a 10$\times$10$\times$1 $k$-mesh during the self-consistency cycle.

We first determine the valence band maximum (VBM) for monolayer WSe$_2$, as well as the VBM for the three stacking configurations, all referenced to the vacuum level. Next, the VBM of three stackings are adjusted by subtracting the VBM of monolayer WSe$_2$ for simplicity. The results are presented in Table \ref{table-1} and are subsequently used to fit the parameters of the continuum model. Figure \ref{fig7} illustrates the electron density distribution for the split VBM in the three stacking configurations, which, together with the VBM energies, uniquely determines the fitting parameters $(t, V, \phi)$ in the continuum model.



\begin{table}[!tb]
    \centering
    \tabcolsep=0.3cm
    \renewcommand\arraystretch{1.3}
    \caption{The energies of split VBM at AA, AB and AC stackings. The energy unit is meV.} \label{table-1}
    \begin{tabular}{c|ccc}
        \hline\hline
         \diagbox{VBM}{Stacking} & AA & AB & AC \\
        \hline
     $E_{v1}$ &     -32.2        &  -30.1  &  -30.1  \\ 
     \hline
    $E_{v2}$ &     -44.1       &  -103.7  &  -103.7  \\                                     
        \hline\hline
    \end{tabular}
 \end{table}

\begin{figure}[tbp] \centering
	\includegraphics[width=0.99\linewidth]{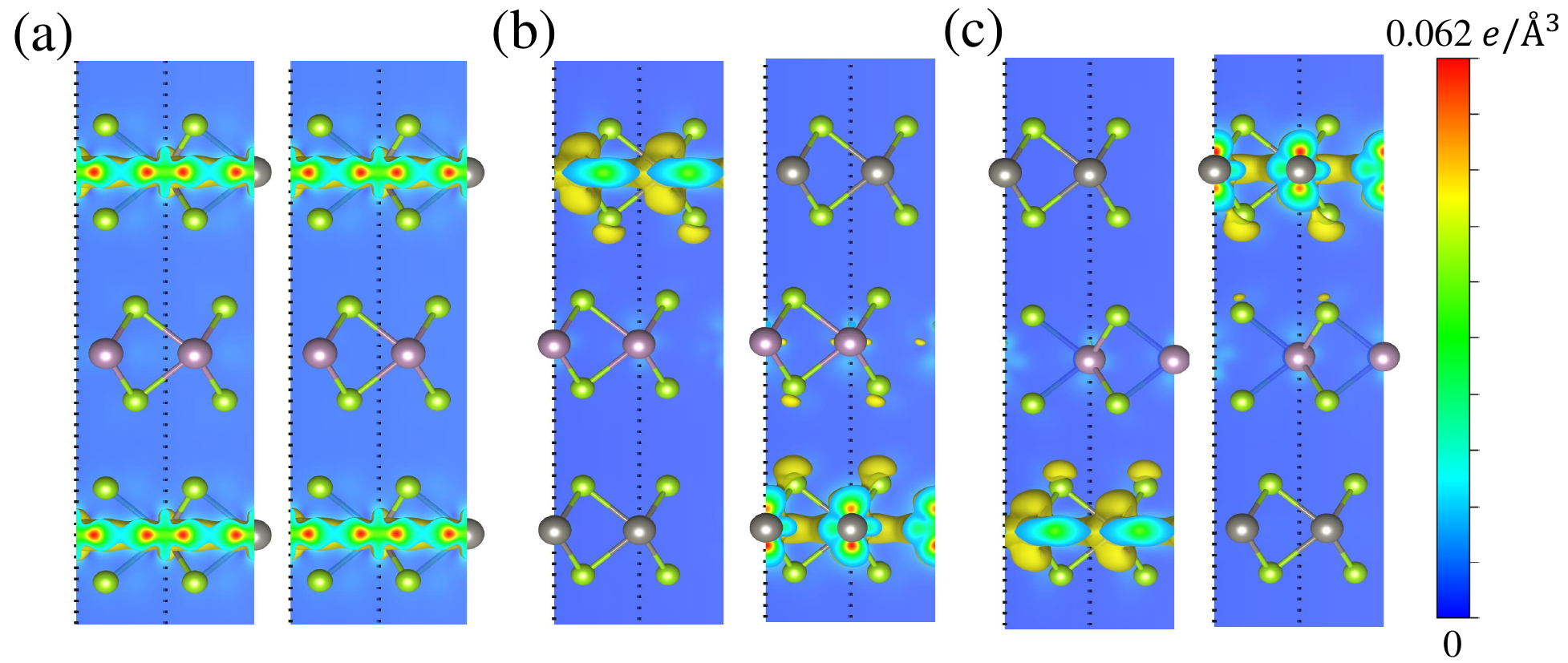}
	\caption{(a)-(c) show the electron density distribution for the lower valence band (left panel) and the higher valence band (right panel) in AA, AB, and AC stacking configurations, respectively. }
	\label{fig7}
\end{figure}

\section{SYMMETRY ANALYSIS OF THE BLOCH STATES}
\label{app:symmetry}
The Bloch wave function $\ket{\Psi_{n,\boldsymbol{Q}}}$ at the high-symmetry points $\boldsymbol{\kappa}$, $\boldsymbol{\kappa}^\prime$ and $\boldsymbol{\gamma}$ is symmetric under the operation $\hat{C}_3$, i.e., 
\begin{equation} \label{eigeneqs}
  \begin{split}
    \hat{C}_3\ket{\Psi_{n,\boldsymbol{Q}}}=e^{i2\pi {L}_{n,\boldsymbol{Q}}/3}\ket{\Psi_{n,\boldsymbol{Q}}},
  \end{split}
\end{equation}
where $e^{i 2\pi {L}_{n,\boldsymbol{Q}}/3}$ is the eigenvalue associated with the eigenstate $\ket{\Psi_{n,\boldsymbol{Q}}}$ and the angular momentum is denoted by ${L}_{n,\boldsymbol{Q}}$.
The angular momentum of the Bloch state $\ket{\Psi_{n,\boldsymbol{Q}}}$ for different topological phases is summarized in Table \ref{table-2}.
\begin{table}[!tb]
    \centering
    \tabcolsep=0.3cm
    \renewcommand\arraystretch{1.3}
    \caption{Angular momentum of Bloch state $\ket{\Psi_{n,\boldsymbol{Q}}}$ at high-symmetry momenta in phases I and II.} \label{table-2}
    \begin{tabular}{c|cccc}
        \hline\hline
        Phase  & \diagbox{$\boldsymbol{Q}$}{$L_{n,\boldsymbol{Q}}$}{states} & $\ket{\Psi_{1,\boldsymbol{Q}}}$ &$\ket{\Psi_{2,\boldsymbol{Q}}}$ &$\ket{\Psi_{3,\boldsymbol{Q}}}$\\
        \hline
                                             & $\boldsymbol{\gamma}$ & 0 & -1 & 0 \\
$\rm{\uppercase\expandafter{\romannumeral1}}$& $\boldsymbol{\kappa}$ & 0 & -1 & 0\\
                                             & $\boldsymbol{\kappa^\prime}$ & 1 & 1 & 0 \\
        \hline                                     
                                             & $\boldsymbol{\gamma}$ & 0 & -1 & 0 \\
$\rm{\uppercase\expandafter{\romannumeral2}}$& $\boldsymbol{\kappa}$ & 0 & -1 & 0\\
                                             & $\boldsymbol{\kappa^\prime}$ & 1 & 0 & 1 \\                      
        \hline\hline
    \end{tabular}
 \end{table}
Based on the values of $L_{2,\boldsymbol{\kappa}^\prime}$ and $L_{3,\boldsymbol{\kappa}^\prime}$ in Phase $\rm{\uppercase\expandafter{\romannumeral1}}$ and Phase $\rm{\uppercase\expandafter{\romannumeral2}}$, it can be determined that a band inversion occurs between bands $n=2$ and $n=3$ at $\boldsymbol{\kappa^\prime}$.


\section{GUTZWILLER METHOD}
\label{app:Gutzwiller}
Within the mean-field theory, we use the bosonic Gutwziller method \cite{jaksch1998cold,krauth1992gutzwiller,lanata2012efficient,zeng2024} to investigate the ground state phases of IXs.
We start from the trial ground state wave function for the three orbitals, 
\begin{equation} \label{Gutzwiller}
    \ket{GW} = \prod_{\alpha}\ket{\phi_{1_\alpha}} \otimes \cdots \otimes \ket{\phi_{N_\alpha}},
\end{equation}
where $\alpha=\{A,B,O\}$ labels the Wannier orbital, $N_\alpha$ is the total number of the lattice sites associated with Wannier orbital $\alpha$, and the local wave function at site $i$ of the orbital $\alpha$ is given by $\ket{\phi_{i_\alpha}}=\sum^{\infty}_{n=0}c^{i_\alpha}_{n}\ket{n}_{i_\alpha}$.
Here, $c^{i_\alpha}_{n}$ is the variational parameter and $\ket{n}_{i_\alpha}$ represents the corresponding local Fock state.
By substituting the variational wave function into the total energy of the EMBH model $E=\bra{GW}\hat{H}\ket{GW}$, one can obtain the expression of energy functional, i.e.,
\begin{align} \label{mean-field energy}
E ({c^{i_\alpha}_{n}}) =&\sum_{\alpha,i_{\alpha}}\left[\frac{U_{\alpha}}{2}\sum_{n=0}^{\infty}\left|c_{n}^{\left(i_{\alpha}\right)}\right|^{2}n(n-1)-t^{(0)}_{i_\alpha,i_\alpha}\sum_{n=0}^{\infty}\left|c_{n}^{\left(i_{\alpha}\right)}\right|^{2}n\right.\nonumber\\
&+\sum_{j_{\alpha^\prime}}\frac{V_{\alpha,\alpha^\prime}}{2}\left(\sum_{n=0}^{\infty}\left|c_{n}^{\left(i_{\alpha}\right)}\right|^{2}n\right)\left(\sum_{n=0}^{\infty}\left|c_{n}^{\left(j_{\alpha^\prime}\right)}\right|^{2}n\right)\nonumber\\
&+\sum_{j_\alpha}\frac{V_{\alpha,\alpha}}{2}\left(\sum_{n=0}^{\infty}\left|c_{n}^{\left(i_{\alpha}\right)}\right|^{2}n\right)\left(\sum_{n=0}^{\infty}\left|c_{n}^{\left(j_{\alpha}\right)}\right|^{2}n\right)\nonumber\\
&-\sum_{j_{\alpha^\prime}}\frac{t^{(1)}_{i_\alpha,j_{\alpha^\prime}}}{2}\left(\sum_{n=0}^{\infty}\sqrt{n+1}c_{n+1}^{*(i_\alpha)}c_{n}^{\left(i_\alpha\right)}\right)\nonumber\\
&\times\left(\sum_{n=0}^{\infty}\sqrt{n+1}c_{n}^{*\left(j^\prime_\alpha\right)}c_{n+1}^{\left(j^\prime_\alpha\right)}\right)\nonumber\\
&-\sum_{j_{\alpha^\prime}}\frac{t^{*(1)}_{i_\alpha,j_{\alpha^\prime}}}{2}\left(\sum_{n=0}^{\infty}\sqrt{n+1}c_{n+1}^{*(j^\prime_\alpha)}c_{n}^{\left(j^\prime_\alpha\right)}\right)\nonumber\\
&\times\left(\sum_{n=0}^{\infty}\sqrt{n+1}c_{n}^{*\left(i_\alpha\right)}c_{n+1}^{\left(i_\alpha\right)}\right)\nonumber\\
&-\sum_{j_{\alpha}}\frac{t^{(2)}_{i_\alpha,j_\alpha}}{2}\left(\sum_{n=0}^{\infty}\sqrt{n+1}c_{n+1}^{*(i_\alpha)}c_{n}^{\left(i_\alpha\right)}\right)\nonumber\\
&\times\left(\sum_{n=0}^{\infty}\sqrt{n+1}c_{n}^{*\left(j_\alpha\right)}c_{n+1}^{\left(j_\alpha\right)}\right)\nonumber\\
&-\sum_{j_{\alpha}}\frac{t^{*(2)}_{i_\alpha,j_\alpha}}{2}\left(\sum_{n=0}^{\infty}\sqrt{n+1}c_{n+1}^{*(j_\alpha)}c_{n}^{\left(j_\alpha\right)}\right)\nonumber\\
&\left.\times\left(\sum_{n=0}^{\infty}\sqrt{n+1}c_{n}^{*\left(i_\alpha\right)}c_{n+1}^{\left(i_\alpha\right)}\right)\right].
\end{align}
The ground state properties of the EMBH model is determined by globally minimizing Eq. (\ref{mean-field energy}) with respect to the variational parameters. In practice, for simplicity, we have assumed that the ground state has the moir\'e lattice periodicity, i.e., the ground state wave function varies only within a moir\'e unit cell.

\bibliographystyle{apsrev4-2}
\end{document}